\documentclass[pra,twocolumn]{revtex4-1}
\usepackage{float,graphicx,amsmath,array}

\begin{document}
\title{ Experimental Demonstration of Non-Destructive Discrimination of Arbitrary Set of Orthogonal Quantum States Using 5-qubit IBM Quantum Computer on Cloud}
\author{Ayan Majumder}
\email{ms14103@iisermohali.ac.in}
\affiliation{Department of Physical Sciences,
Indian
Institute of Science Education and
Research, Mohali-140306, India.}

\author{Anil Kumar}
\email{anilnmr@iisc.ac.in}
\affiliation{
Department of Physics, NMR Research Centre, Indian Institute of Science, Bangalore-560012, India.}
\begin{abstract} A protocol for non-destructive descrimination of arbitrary set of orthogonal quantum states was proposed by V. S. Manu et al., using an  algorithm based on quantum phase estimation. IBM Corporation has released a superconductivity based 5-qubit (5-qubit transmon bowtie chip 3 and IBM 5-qubit real processor) quantum computer named “Quantum Experience” and placed it on “cloud”. In this paper we take advantage of the online availability of those real quantum processors(ibmqx2 amd ibmqx4) and carry out the above protocol that has experimentally demonstrated earlier using NMR quantum processor. Here, we set up experiments for arbitrary one-qubit and two-qubit orthogonal quantum states. The experiment confirmed that the arbitrary orthogonal quantum states can be discriminate in a nondestructive manner with a high fidelity. We compare the outcomes of those experiments which are done by ibmqx2 and ibmqx4 processors. Here, we also show the state tomography for the single qubit experiments.
\end{abstract}
\maketitle

\section{Introduction}
\label{intro}
There are many protocols\cite{Walgate,Ghosh,Virmani,Chen} available for
orthogonal quantum state discrimination. Phase estimation plays an
important role in quantum computation and
is a key element of many quantum algorithms\cite{Majumder,Shi}. When
the phase estimation is combined with other quantum algorithms, it can be employed to perform certain computational tasks such as quantum counting, order finding
and factorization. By defining an operator with preferred eigen-values,
phase estimation can be used for discrimination of quantum states with certainty. It preserves
the state since local operations on ancilla qubit measurements do not affect the quantum state.
Besides this, several groups use the phase estimation algorithm in quantum chemistry\cite{Du,Guzik}, genetics\cite{Beltra} and also in quantum cryptography\cite{Majumder,Shi}.
IBM Corporation has released the Quantum Experience
which allows users to access 5-qubit quantum processors(ibmqx2 qnd ibmqx4). 
We take advantage of the online availability of this 
real hardware and present the non-destructive discrimination of orthogonal quantum states. 
Here, we experimentally
implement this protocol which is given by V. S. Manu
et al.\cite{Manu}, using the five-qubit superconductivity based
quantum computer. A comparison of the outcomes of those experiments 
using IBM quantum processors and the outcomes which are obtained earlier in the
 NMR quantum processor. IBM quantum processor
which we have used here, is placed at T.J.Watson lab,
York Town, USA. Till now several groups have used
the 5-qubit quantum computer to demonstrate various
experiments using ibmqx2 quantum processor\cite{Mitali,Kalra,Behera,Ma,Devitt}.  Recently
several groups have discussed hardware for superconductivity based quantum processor\cite{Mandip,You,Clarke,Rosenberg}. 
\section{Theory}
\label{theory}
\begin{figure}[H]
\includegraphics[scale=.45]{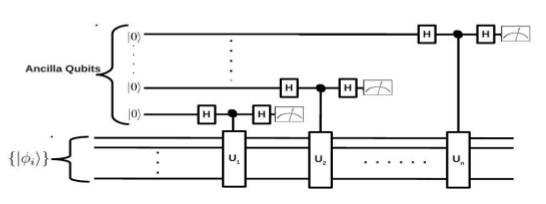}
\caption{The general circuit for non-destructive discrimination of orthogonal quantum state . $n$ ancilla qubits are prepared in the state $|00...0\rangle$. Here $H$ represents
Hadamard quantum gate, and the meter represents a measurement of the qubit state in $\sigma_z$ basis.}
\label{n qubit case}
\end{figure}

According to reference\cite{Manu}
for $n$ qubit quantum states the Hilbert space dimension is $2^n$,
means there are $2^n$ independent orthogonal quantum states. So we have
to design a quantum circuit to discrimininate
a set of $2^n$ orthogonal quantum states. Consider a set
of $2^n$ orthogonal states $\{\phi_i\}$, where $i=1, 2, ....,2^n$.
We need $n$ ancilla qubits for proper discrimination of $2^n$ orthogonal quantum states.
Besides this $n$ ancilla qubits, the discrimination circuit requires $n$ Controlled
Operations. Selecting these $n$ operators $\{U_j\}$ (where
$j=1, 2, ...,n$) is the main task in designing the algorithm.
The set of $\{U_j\}$ depends on the $2^n$ orthogonal states in
such a way that the set of orthogonal vectors forms the
eigen-vector set of the operators, with eigen-values $\pm1$.
The sequence of $+1$ and $-1$ in the eigen-values should
be defined in a special way, as following. Let $\{e^i_j\}$
(where $i=1, 2,...,2^n$) be the eigen-value array of $U_j$, and it
should satisfy following conditions,
\begin{itemize}
\item{Eigen-value arrays $\{e^i_j\}$ of all operators
$\{U_j\}$ should contain equal number of $+1$ and $-1$.}
\item{For the first operator $U_1$, the eigen-value
array $\{e^i_1\}$ can be any possible sequence of $+1$ and $-1$
with condition-1.}
\item{The restriction on eigen-value arrays
starts from $U_{j=2}$ onwards. The eigen-value array $\{e^i_2\}$
of operator $U_2$ should not be equal to $\{e^i_1\}$ or its
complement and satisfying the condition-1.}
\item{By generalizing the condition-3, the
eigen-value array $\{e^i_k\}$ of operator $U_k$ should not be
equal to $\{e^i_m\}$ ($m = 1, 2, ...,k-1$) or its complement.}
\end{itemize}
Let $M_j$ be the diagonal matrix formed by eigen-value
array $\{e^i_j\}$ of $U_j$ operator. The operator $U_j$ is directly related to
$M_j$ by a unitary transformation given by Eq.\ref{eq1},
\begin{equation} 
U_j = V^{-1} \times M_j \times V, 
\label{eq1}
\end{equation}
where $V$ is the matrix formed by the column vectors
$\{|\phi_i\rangle\}, V = [ |\phi_1\rangle |\phi_2\rangle |\phi_3\rangle ..... |\phi_n\rangle]$ and $V^{-1}$ is the inverse matrix of $V$.
The general circuit diagram for $n$-qubit orthogonal qauntum states discrimination is shown in FIG.\ref{n qubit case}.
\section{single qubit orthogonal quantum states discrimination}
\label{single}
\subsection{Single Qubit Orthogonal Quantum States Discrimination Using ibmqx4}
\begin{figure}[H]
\includegraphics[scale=.30]{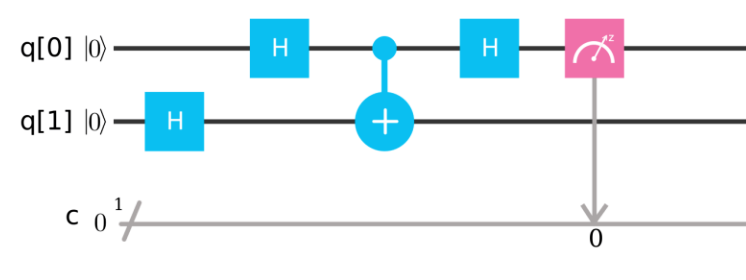}
\caption{Simulation-1: The circuit for non-destructive discrimination of single qubit orthogonal quantum states(designed in IBM quantum simulator). Here, q[0] represents the ancilla qubit which is in state  $|0\rangle$ and q[1] represents the quantum state $|\phi_1\rangle=\frac{1}{\sqrt{2}}(|0\rangle+|1\rangle)$(Initially, q[1] is in state $|0\rangle$, then we apply the $H$-gate on this state to get state $|\phi_1\rangle$). The meter represents a measurement of the qubit state in $\sigma_z$ basis.}
\label{single qubit case1}
\end{figure}

\begin{figure}[H]
\includegraphics[scale=.40]{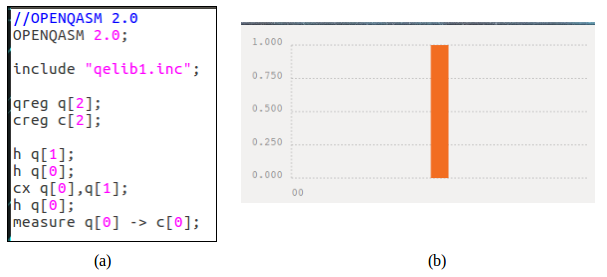}
\caption{(a)Showes the code for the simulation of discrimination of $|\phi_1\rangle=\frac{1}{\sqrt{2}}(|0\rangle+|1\rangle)$ state.(b)Showes the simulation result of this experiment. Here classical bit 0(cbit0) is the simulation result.}
\label{code1}
\end{figure}

\begin{figure}[H]
\includegraphics[scale=.30]{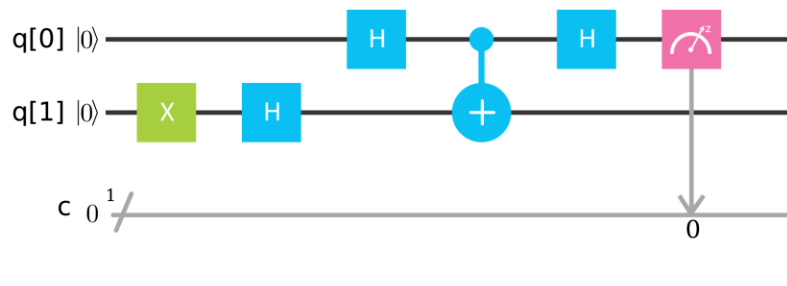}
\caption{Simulation-2: The circuit for non-destructive discrimination of single qubit orthogonal quantum states(designed in IBM quantum simulator). Here, q[0] represents the ancilla qubit which is in state  $|0\rangle$ and q[1] represents the quantum state $|\phi_2\rangle=\frac{1}{\sqrt{2}}(|0\rangle-|1\rangle)$(Initially, q[1] is in state $|0\rangle$, then we apply the $\sigma_x$ and $H$-gate respectively on this state to get state $|\phi_2\rangle$). The meter represents a measurement of the qubit state in $\sigma_z$ basis.}
\label{single qubit case2}
\end{figure}

\begin{figure}[H]
\includegraphics[scale=.40]{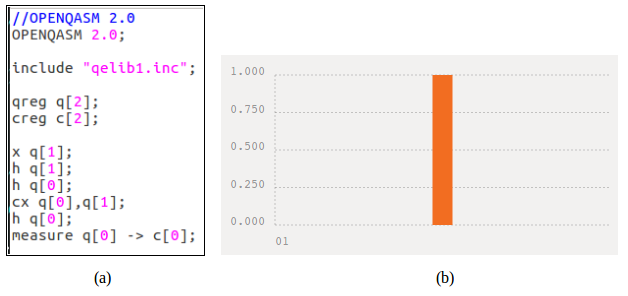}
\caption{(a)Showes the code for the simulation of discrimination of $|\phi_1\rangle=\frac{1}{\sqrt{2}}(|0\rangle-|1\rangle)$ state.(b)Showes the simulation result of this experiment. Here classical bit 0(cbit0) is the simulation result.}
\label{code2}
\end{figure}

For a single qubit orthogonal quantum states, the Hilbert space dimension
is $n=2$. According to the theory(Sec.\ref{theory}) we can discriminate a state from a set of two orthogonal quantum states. Let, the set of single qubit orthogonal quantum states 
is $ \{|\phi_1\rangle=\frac{1}{\sqrt{2}}(|0\rangle+|1\rangle),|\phi_2\rangle=\frac{1}{\sqrt{2}}
(|0\rangle-|1\rangle)\}$. The quantum state discrimination circuit can be designed 
by the general procedure which is 
discussed in Sec.\ref{theory}. The $V$ matrix for the states
$\{|\phi_1\rangle, |\phi_2\rangle\}$ is given by Eq.\ref{eq2},
\[V=[|\phi_1\rangle \ |\phi_2\rangle]\]
where,
\[|0\rangle=\begin{bmatrix} 1 \\ 0 \end{bmatrix} \ , \ |1\rangle=\begin{bmatrix} 0 \\ 1 \end{bmatrix}\]
\begin{equation} 
V=\frac{1}{\sqrt{2}}\begin{bmatrix} 1&1 \\ 1&-1 \end{bmatrix} 
\label{eq2}
\end{equation}
\[V^{-1}=\frac{Adj[V]}{|V|}\]
where,
\[Adj[V]=\frac{1}{\sqrt{2}}\begin{bmatrix} -1&-1 \\ -1&1 \end{bmatrix} \ , \ |V|=-1\]
\[V^{-1}=\frac{1}{\sqrt{2}}\begin{bmatrix} 1&1 \\ 1&-1 \end{bmatrix}\]
According to the theory(Sec.\ref{theory}) $M$-operator can be either
\begin{equation} 
M=\begin{bmatrix} 1&0 \\ 0&-1 \end{bmatrix} or \begin{bmatrix} -1&0 \\ 0&1 \end{bmatrix}.
\label{eq3}
\end{equation}
Here, we take $M=\begin{bmatrix} 1&0 \\ 0&-1 \end{bmatrix}$.
According to Eq.\ref{eq1},
\begin{equation} 
U = V^{-1} \times M \times V = \begin{bmatrix} 0&1 \\ 1&0 \end{bmatrix} = \sigma_x .
\label{eq4}
\end{equation} 
Therefore, the \textit{Controllrd-U} operation\cite{Chaung,Cross} is \textit{Controlled-NOT}(CNOT) operation for this case.
Before the experiment, we have done a simulation in Custom Topology(IBM quantum simulator) to verify our circuit. The simulation for the single qubit case is shown in FIG.\ref{single qubit case1} and FIG.\ref{single qubit case2}, having only one state and one ancilla qubit. We have also shown the simulation results and codes in FIGs.\ref{code1},\ref{code2} .


\begin{figure}[H]
\includegraphics[scale=.6]{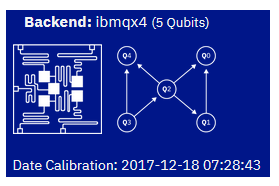}
\caption{Schematic circuit diagram of ibmqx4 processor(transmon bowtie chip 3). The connectivity is provided by two coplanar waveguide (CPW) resonators with resonances around 6.6 GHz (coupling Q2, Q3 and Q4) and 7.0 GHz (coupling Q0, Q1 and Q2). Each qubit has a dedicated CPW for control and readout.
Coupling map $= {1: [0], 2: [0, 1, 4], 3: [2, 4]}$ where, a: [b] means a CNOT with qubit a 
as control and b as target can be implemented.}
\label{transmon bowtie chip}
\end{figure}

\begin{figure}[H]
\includegraphics[scale=.22]{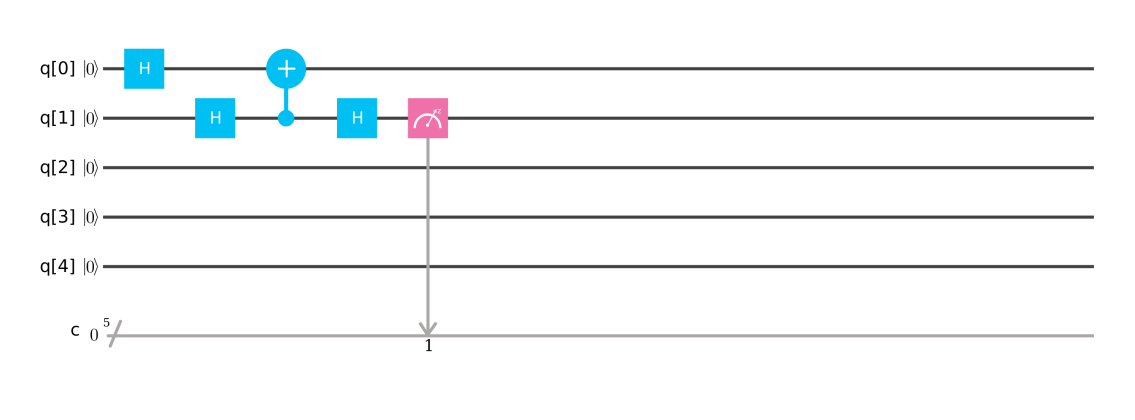}
\caption{Experiment-1: The circuit for non-destructive discrimination of single qubit orthogonal quantum states(designed in 5-qubit transmon bowtie chip 3 ). Here, q[1] represents the ancilla qubit which is in state  $|0\rangle$ and q[0] represents the quantum state $|\phi_2\rangle=\frac{1}{\sqrt{2}}(|0\rangle+|1\rangle)$(Initially, q[0] is in state $|0\rangle$, then we apply the $H$-gate on this state to get state $|\phi_2\rangle$). The meter represents a measurement of the qubit state in $\sigma_z$ basis.}
\label{single qubit case3}
\end{figure}

\begin{figure}[H]
\includegraphics[scale=.70]{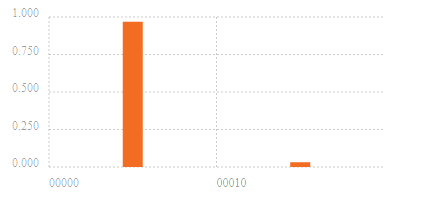}
\caption{The outcomes of the experiment(FIG.\ref{single qubit case3}) in $\sigma_z$ basis. Here, classical bit 1(cbit1) is the result of the experiment. Probabilities of finding the ancilla qubit in state $|0\rangle$ and $|1\rangle$ are 0.969 and 0.031 respectively. The above result is obtained by taking 8192 number of shots.}
\label{single qubit case3a}
\end{figure}

\begin{figure}[H]
\includegraphics[scale=.22]{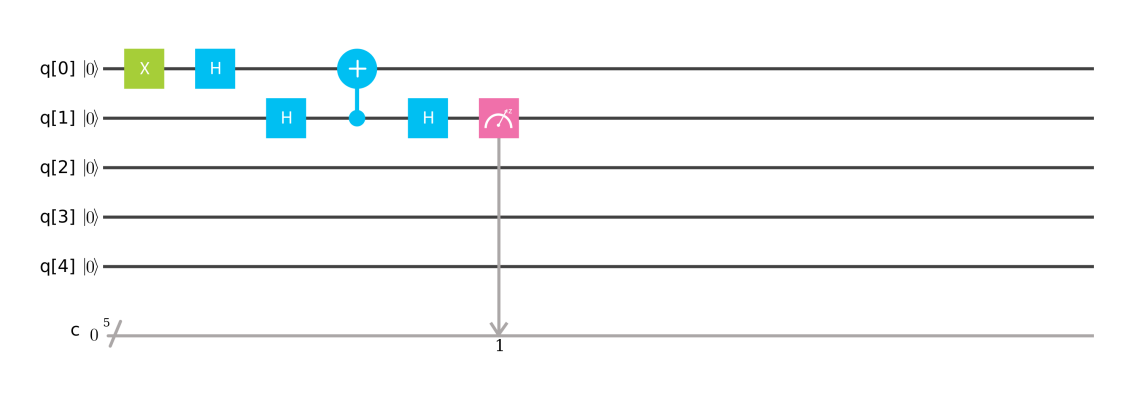}
\caption{Experiment-2: The circuit for non-destructive discrimination of single qubit orthogonal quantum states(designed in 5-qubit transmon bowtie chip 3). Here, q[1] represents the ancilla qubit which is in state  $|0\rangle$ and q[0] represents the quantum state $|\phi_2\rangle=\frac{1}{\sqrt{2}}(|0\rangle-|1\rangle)$(Initially, q[0] is in state $|0\rangle$, then we apply the $\sigma_x$ and $H$-gate respectively on this state to get state $|\phi_2\rangle$). The meter represents a measurement of the qubit state in $\sigma_z$ basis.}
\label{single qubit case4}
\end{figure}

\begin{figure}[H]
\includegraphics[scale=.5]{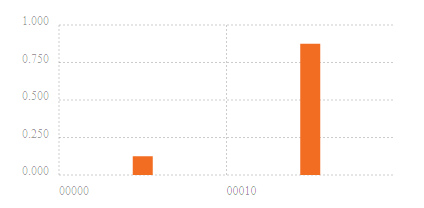}
\caption{The outcomes of the experiment(FIG.\ref{single qubit case4}) in $\sigma_z$ basis. Here, classical bit 1(cbit1) is the result of the experiment. Probabilities of finding the ancilla qubit in state $|1\rangle$ and $|0\rangle$ are 0.875 and 0.125 respectively. The above result is obtained by taking 8192 number of shots.}
\label{single qubit case4a}
\end{figure}

\begin{figure}[H]
\includegraphics[scale=.41]{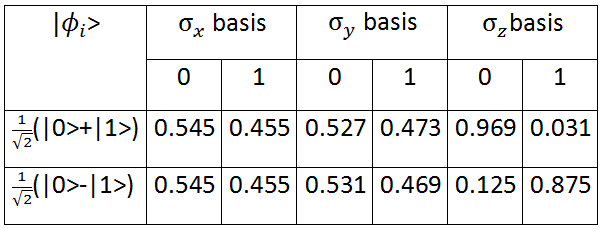}
\caption{The outcomes of the experiments(FIG.\ref{single qubit case3}and FIG.\ref{single qubit case4}) in $\sigma_x, \sigma_y$ and $\sigma_z$ basis. Probabilities of finding the ancilla qubit in state $|1\rangle$ and $|0\rangle$ are shown in this table. The above result is obtained by taking 8192 number of shots.}
\label{T1}
\end{figure}

The real experimental circuits which are shown in FIG.\ref{single qubit case3} and FIG.\ref{single qubit case4}, is designed in 5-qubit transmon bowtie chip(ibmqx4 quantum processor) on cloud. The chip architecture which is important for designing a quantum circuit for an experiment, is shown in FIG.\ref{transmon bowtie chip}. According to this coupling map of the ibmqx4 processor, we consider q[0] as a state qubit and q[1] as a ancilla qubit. The results of those experiments are given in FIGs \ref{single qubit case3a},\ref{single qubit case4a} and \ref{T1}.


\begin{figure}[H]
\includegraphics[scale=.5]{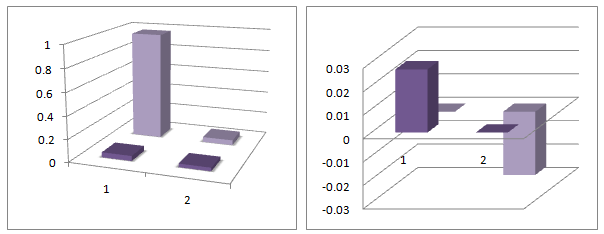}
\caption{The graphical representation of experimental density matix(Eq.\ref{eq9}). The left part shows the real part of the experimental density matrix and right part shows the imaginary part of the density matrix.}
\label{tomography1}
\end{figure}

\begin{figure}[H]
\includegraphics[scale=.65]{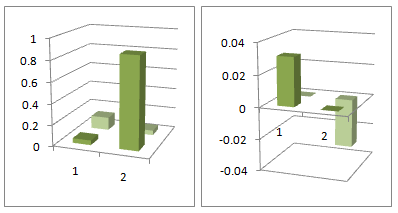}
\caption{The graphical representation of experimental density matix(Eq.\ref{eq11}). The left part shows the real part of the experimental density matrix and right part shows the imaginary part of the density matrix.}
\label{tomography2}
\end{figure}

\begin{figure}[H]
\includegraphics[scale=.35]{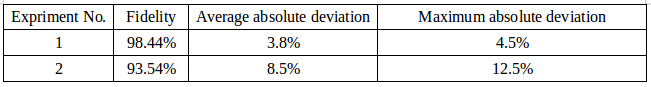}
\caption{Results for the experiment-1 and experiment-2.}
\label{result1}
\end{figure}

Here, we calculate fidelity(Eq.\ref{eq5}), average absolute deviation($\langle\Delta x \rangle$)(Eq.\ref{eq6}) and maximum absolute
deviation($\Delta x_{max}$)(Eq.\ref{eq7}) of those experiments. Results are shown in FIG.\ref{result1} and the state tomography\cite{Mitali} of the experimental desity matrix for both experiments are shown in FIGs.\ref{tomography1},\ref{tomography2}.
\begin{equation} 
Fidelity=Tr{\sqrt{\sqrt{\rho^T}.{\rho^E}.{\sqrt{\rho^T}}}}
\label{eq5}
\end{equation}
In Eq.\ref{eq5}, $\rho^T$ is the theoretical density matrix and
$\rho^E$ is the experimental density matrix.
\begin{equation} 
\langle\Delta x \rangle = \frac{1}{N^2}\sum_{i,j=1}^{N} |x^T_{i,j}-x^E_{i,j}|
\label{eq6}
\end{equation}
\begin{equation} 
\Delta x_{max} = Max|x^T_{i,j}-x^E_{i,j}|
\label{eq7}
\end{equation}
In Eq.\ref{eq6} and Eq.\ref{eq7} $x^T_{i,j}$ is the element of theoretical
density matrix and $x^E_{i,j}$ is the element of experimental
density matrix.
According to the experiment which is shown in FIG.\ref{single qubit case3}, the theoretical density matix($\rho^T$) of the ancilla qubit is given by Eq.\ref{eq8}.
\begin{equation} 
\rho^T_{|0\rangle}=|0\rangle\langle 0|=\begin{bmatrix} 1&0 \\ 0&0 \end{bmatrix}
\label{eq8}
\end{equation}
From the measurement results(FIG.\ref{T1}), we can construct the experimental density matrix($\rho^E$) which is given by the Eq.\ref{eq9}.
\[\rho^E_{|0\rangle}=Re[\rho^E_{|0\rangle}]+iIm[\rho^E_{|0\rangle}]\] 
\begin{equation} 
\rho^E_{|0\rangle}=\begin{bmatrix} 0.969&0.045-0.027i \\ 0.045+0.027i&0.031 \end{bmatrix}
\label{eq9}
\end{equation}
In a similar way, we can calculate the theoretical($\rho^T_{|1\rangle}$) and the experimental($\rho^E_{|1\rangle}$) density matrix for the experiment which is shown in FIG.\ref{single qubit case4}. For this experiment, $\rho^T_{|1\rangle}$ and $\rho^E_{|1\rangle}$
are given by Eq.\ref{eq10} and Eq.\ref{eq11} respectively.
\begin{equation} 
\rho^T_{|1\rangle}=|1\rangle\langle 1|=\begin{bmatrix} 0&0 \\ 0&1 \end{bmatrix}
\label{eq10}
\end{equation}
\begin{equation} 
\rho^E_{|1\rangle}=\begin{bmatrix} 0.125&0.045-0.031i \\ 0.045+0.031i&0.875 \end{bmatrix}
\label{eq11}
\end{equation}
For the generalization of single qubit case, we can consider a set of arbitrary single qubit orthogonal quantum states which is $\{|\phi_1\rangle=\alpha|0\rangle + \beta|1\rangle,|\phi_2\rangle=\alpha|0\rangle - \beta|1\rangle \}$. Where, $\alpha$ and $\beta$ are real numbers satisfying, $|\alpha|^2+|\beta|^2=1$. According to the theory(Sec.\ref{theory}), We can construct the $U$ operator\cite{Chaung,Cross} for eignvalue array $\{1,-1\}$(Eq.\ref{eq14}).
\begin{equation} 
U=\begin{bmatrix} Cos(\theta)&Sin(\theta) \\ Sin(\theta)& -Cos(\theta) \end{bmatrix}
\label{eq14}
\end{equation}
Where, 
\[\theta = 2 \times {Tan}^{-1}(\frac{\beta}{\alpha})\]
\subsection{Single Qubit Orthogonal Quantum States Discrimination Using ibmqx2}
\begin{figure}[H]
\includegraphics[scale=.5]{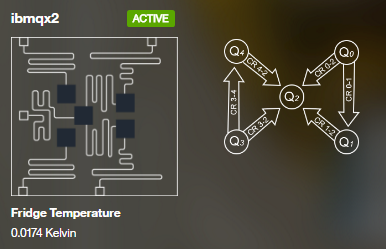}
\caption{Schematic circuit diagram of ibmqx2 processor(5-qubit IBM real quantum processor). The connectivity is provided by two coplanar waveguide (CPW) resonators with resonances around 6.0 GHz (coupling Q2, Q3 and Q4) and 6.5 GHz (coupling Q0, Q1 and Q2). Each qubit has a dedicated CPW for control and readout. 
Coupling map $= {0: [1, 2], 1: [2], 3: [2, 4], 4: [2]}$ where, a: [b] means a CNOT with qubit a 
as control and b as target can be implemented.}
\label{ibmqx2}
\end{figure}

\begin{figure}[H]
\includegraphics[scale=.22]{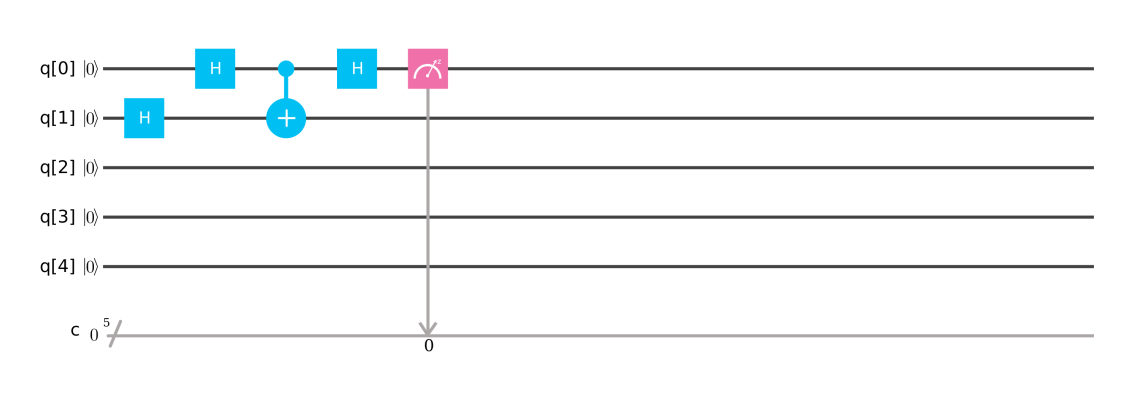}
\caption{Experiment-3: The circuit for non-destructive discrimination of single qubit orthogonal quantum states(designed in 5-qubit real quantum processor). Here, q[0] represents the ancilla qubit which is in state  $|0\rangle$ and q[1] represents the quantum state $|\phi_2\rangle=\frac{1}{\sqrt{2}}(|0\rangle+|1\rangle)$(Initially, q[1] is in state $|0\rangle$, then we apply the $H$-gate on this state to get state $|\phi_2\rangle$). The meter represents a measurement of the qubit state in $\sigma_z$ basis.}
\label{ibmqx2a}
\end{figure}

\begin{figure}[H]
\includegraphics[scale=.22]{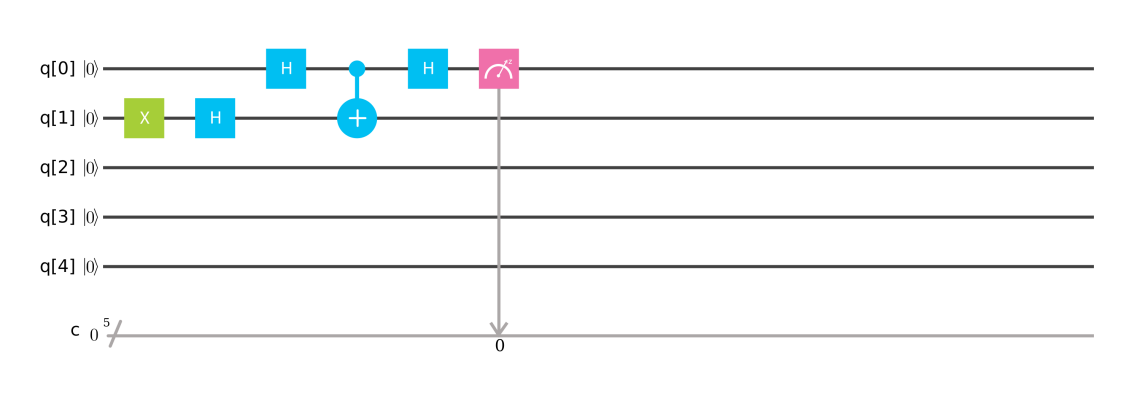}
\caption{Experiment-4: The circuit for non-destructive discrimination of single qubit orthogonal quantum states(designed in 5-qubit real quantum processor). Here, q[0] represents the ancilla qubit which is in state  $|0\rangle$ and q[1] represents the quantum state $|\phi_2\rangle=\frac{1}{\sqrt{2}}(|0\rangle-|1\rangle)$(Initially, q[1] is in state $|0\rangle$, then we apply the $\sigma_x$ and $H$-gate respectively on this state to get state $|\phi_2\rangle$). The meter represents a measurement of the qubit state in $\sigma_z$ basis.}
\label{ibmqx2b}
\end{figure}

\begin{figure}[H]
\includegraphics[scale=.35]{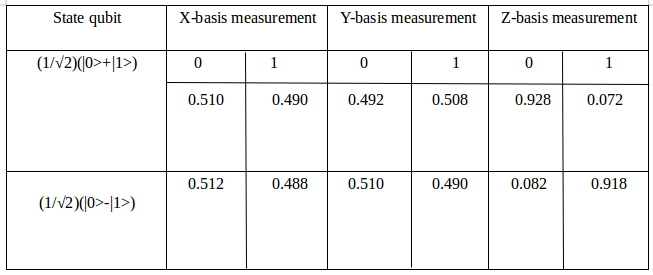}
\caption{The outcomes of the experiments(FIG.\ref{single qubit case3}and FIG.\ref{single qubit case4}) in $\sigma_x, \sigma_y$ and $\sigma_z$ basis. Probabilities of finding the ancilla qubit in state $|1\rangle$ and $|0\rangle$ are shown in this table. The above result is obtained by taking 8192 number of shots.}
\label{T2}
\end{figure}

The real experimental circuits which are shown in FIG.\ref{ibmqx2a} and FIG.\ref{ibmqx2b}, is designed in 5-qubit real quantum processor(ibmqx2) on cloud. The chip architecture which is important for designing a quantum circuit for an experiment, is shown in FIG.\ref{ibmqx2}. According to this coupling map of the ibmqx2 processor, we consider q[1] as a state qubit and q[0] as a ancilla qubit. The results of those experiments are given in FIG \ref{T2}.

\begin{figure}[H]
\includegraphics[scale=.4]{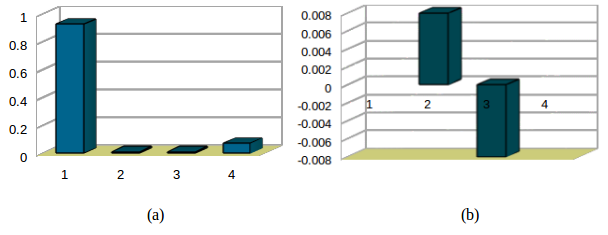}
\caption{The graphical representation of experimental density matix(Eq.\ref{eq12}). The left part shows the real part of the experimental density matrix and right part shows the imaginary part of the density matrix.}
\label{tomography3}
\end{figure}

\begin{figure}[H]
\includegraphics[scale=.4]{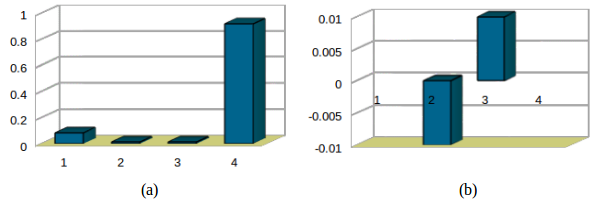}
\caption{The graphical representation of experimental density matix(Eq.\ref{eq13}). The left part shows the real part of the experimental density matrix and right part shows the imaginary part of the density matrix.}
\label{tomography4}
\end{figure}

\begin{figure}[H]
\includegraphics[scale=.35]{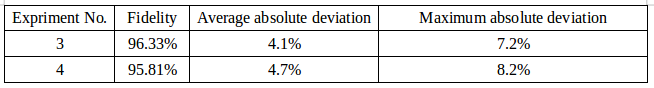}
\caption{The results for the experiment-3 and experiment-4.}
\label{result3}
\end{figure}

Here, we calculate fidelity(Eq.\ref{eq5}), average absolute deviation($\langle\Delta x \rangle$)(Eq.\ref{eq6}) and maximum absolute
deviation($\Delta x_{max}$)(Eq.\ref{eq7}) of those experiments. Results are shown in FIG.\ref{result3}.
From the measurement results(FIG.\ref{T2}), we can construct the experimental density matrix($\rho^E$ for the first experiment (FIG.\ref{ibmqx2a})) which is given by the Eq.\ref{eq12}.
We aslo show the state-tomography\cite{Mitali} of those experiments\ref{tomography3},\ref{tomography4}.
 \begin{equation} 
\rho^E_{|0\rangle}=\begin{bmatrix} 0.928&0.01+0.008i \\ 0.01-0.008i&0.072 \end{bmatrix}
\label{eq12}
\end{equation}
Experimental density matrix for second experiment(FIG.\ref{ibmqx2b}) is given by Eq.\ref{eq13} 
\begin{equation} 
\rho^E_{|0\rangle}=\begin{bmatrix} 0.082&0.012-0.01i \\ 0.012+0.01i&0.918 \end{bmatrix}
\label{eq13}
\end{equation}
\section{two qubit orthogonal quantum states discrimination}
\label{two}
\begin{figure}[H]
\includegraphics[scale=.3]{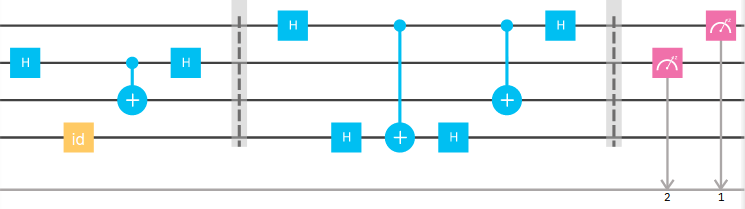}
\caption{Schematic circuit diagram.}
\label{EX1}
\end{figure}

\begin{figure}[H]
\includegraphics[scale=.53]{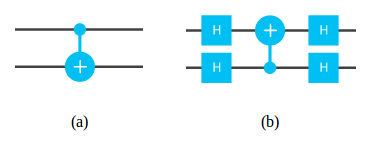}
\caption{(a) Actual CNOT gate. (b) CNOT gate which is used in these experiments.}
\label{CNOT}
\end{figure}

\begin{figure}[H]
\includegraphics[scale=.5]{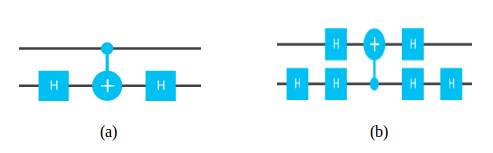}
\caption{(a) Actual CZ gate. (b) CZ gate which is used in these experiments.}
\label{CZ}
\end{figure}

\begin{figure}[H]
\includegraphics[scale=.3]{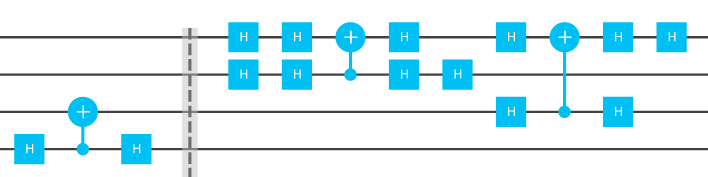}
\caption{Actual circuit diagram which is used in these experiments.}
\label{EX2}
\end{figure}

\begin{figure}[H]
\includegraphics[scale=.2]{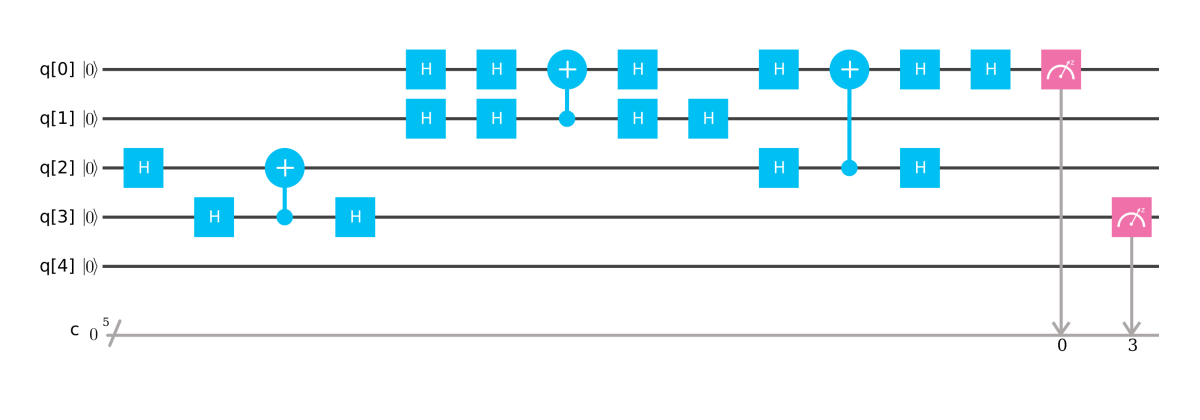}
\caption{Experiment-5: The circuit for non-destructive discrimination of two qubit orthogonal quantum states(designed in ibmqx4). Here, q[0] and q[3] represent the ancilla qubit which are in state  $|00\rangle$. q[1] and q[2] represent the quantum state $|\phi_1\rangle=\frac{1}{\sqrt{2}}(|00\rangle+|01\rangle)$(Initially, q[1] and q[2] are in state $|00\rangle$, then we apply the $H$-gate on the q[2] to get state $|\phi_1\rangle$). The meter represents a measurement of the qubit state in $\sigma_z$ basis.}
\label{ibmqx4a}
\end{figure}

\begin{figure}[H]
\includegraphics[scale=.2]{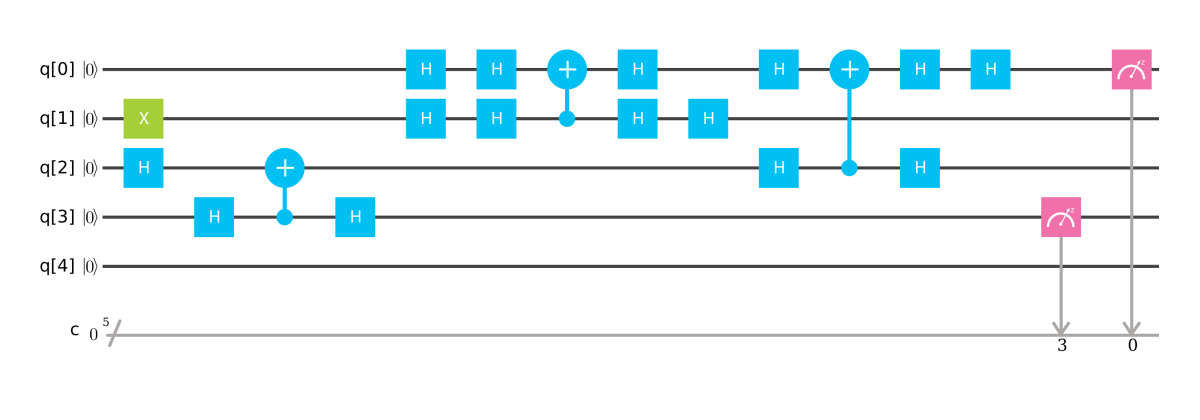}
\caption{Experiment-6: The circuit for non-destructive discrimination of two qubit orthogonal quantum states(designed in ibmqx4). Here, q[0] and q[3] represent the ancilla qubit which are in state  $|00\rangle$. q[1] and q[2] represent the quantum state $|\phi_2\rangle=\frac{1}{\sqrt{2}}(|10\rangle+|11\rangle)$(Initially, q[1] and q[2] are in state $|00\rangle$, then we apply the $\sigma_x$-gate on the qubit q[1] and $H$-gate on the q[2] to get state $|\phi_2\rangle$). The meter represents a measurement of the qubit state in $\sigma_z$ basis.}
\label{ibmqx4b}
\end{figure}

\begin{figure}[H]
\includegraphics[scale=.2]{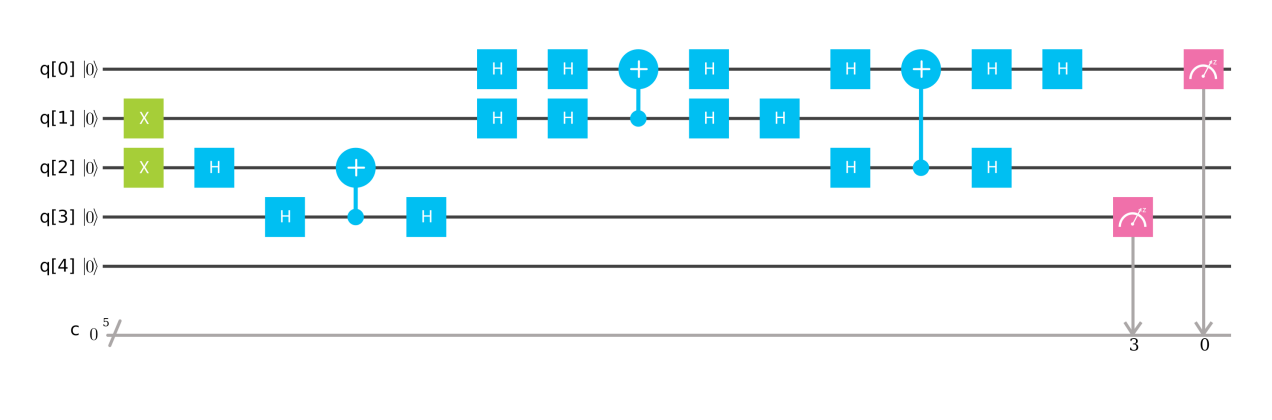}
\caption{Experiment-7: The circuit for non-destructive discrimination of two qubit orthogonal quantum states(designed in ibmqx4). Here, q[0] and q[3] represent the ancilla qubit which are in state  $|00\rangle$. q[1] and q[2] represent the quantum state $|\phi_3\rangle=\frac{1}{\sqrt{2}}(|10\rangle-|11\rangle)$(Initially, q[1] and q[2] are in state $|00\rangle$, then we apply the $\sigma_x$-gate on the qubit q[1] and $\sigma_{x}H$-gate on the q[2] to get state $|\phi_3\rangle$). The meter represents a measurement of the qubit state in $\sigma_z$ basis.}
\label{ibmqx4c}
\end{figure}

\begin{figure}[H]
\includegraphics[scale=.2]{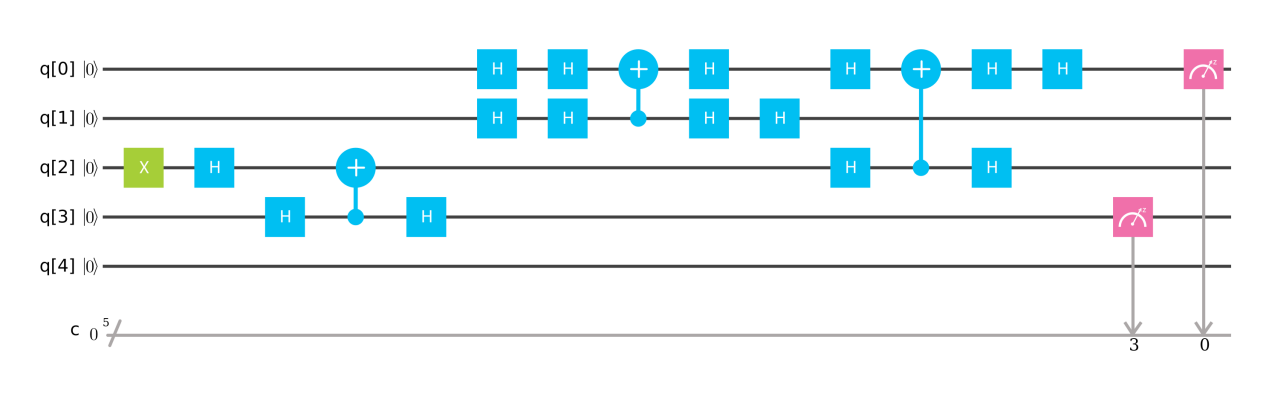}
\caption{Experiment-8: The circuit for non-destructive discrimination of two qubit orthogonal quantum states(designed in ibmqx4). Here, q[0] and q[3] represent the ancilla qubit which are in state  $|00\rangle$. q[1] and q[2] represent the quantum state $|\phi_4\rangle=\frac{1}{\sqrt{2}}(|00\rangle-|01\rangle)$(Initially, q[1] and q[2] are in state $|00\rangle$, then we apply the $\sigma_{x}H$-gate on the q[2] to get state $|\phi_4\rangle$). The meter represents a measurement of the qubit state in $\sigma_z$ basis.}
\label{ibmqx4d}
\end{figure}

In a similar way, we set up four experiments to discriminate the ortogonal two qunit quantum states. The set of two qubit orthogonal quantum state is $\{|\phi_i\rangle\}=\{\frac{1}{\sqrt{2}}(|00\rangle+|01\rangle),\frac{1}{\sqrt{2}}(|10\rangle+|11\rangle),\frac{1}{\sqrt{2}}(|10\rangle-|11\rangle),\frac{1}{\sqrt{2}}(|00\rangle-|01\rangle) \}$.
The $U$ operator\cite{Chaung,Cross} for this case is given by Eqs.\ref{eq15},\ref{eq16}
\begin{equation} 
U_1=\begin{bmatrix} 0&1&0&0 \\ 1&0&0&0 \\ 0&0&0&1 \\ 0&0&1&0 \end{bmatrix} = I \otimes \sigma_{x}
\label{eq15}
\end{equation}
\begin{equation} 
U_1=\begin{bmatrix} 0&1&0&0 \\ 1&0&0&0 \\ 0&0&0&-1 \\ 0&0&-1&0 \end{bmatrix} = \sigma_{z} \otimes \sigma_{x}
\label{eq16}
\end{equation}
Therefore the controlled-$U_1$ operation is controlled-I$\otimes$CNOT and the controlled-$U_2$ operation is controlled-Z$\otimes$CNOT. Here, we prepare the state qubits using Pauli matrices and Hadamard gate. The schematic circuit diagram of the experiment is shown in the FIG.\ref{EX1}. The first part of this figure shows the controlled-$U_1$ operation and second part shows the controlled-$U_2$ operation. In these experiments, we cannot use the controlled-NOT gate directly due to the coupling map of the ibmqx4 processor(FIG.\ref{transmon bowtie chip}). We use it in a different way which is shown in FIG.\ref{CNOT}.
The controlled-$\sigma_z$(CZ) operation are shown in FIG.\ref{CZ}.
The actual circuit diagram of these experiments is shown in FIG.\ref{EX2} . The first part of this figure shows the controlled-$U_1$ operation and second part shows the controlled-$U_2$ operation.
The real quantum circuits are shown in FIGs.\ref{ibmqx4a},\ref{ibmqx4b},\ref{ibmqx4c},\ref{ibmqx4d}.

\

\begin{figure}[H]
\includegraphics[scale=.35]{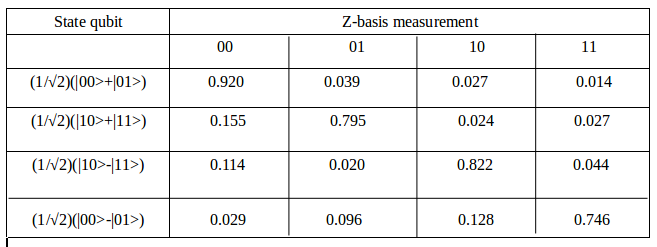}
\caption{The outcomes of the experiments(FIGs.\ref{ibmqx4a},\ref{ibmqx4b},\ref{ibmqx4c} and .\ref{ibmqx4d}) in $\sigma_z$ basis. Probabilities of finding the ancilla qubit in state $|00\rangle, |01\rangle, |10\rangle$ and $|11\rangle$ are shown in this table. The above result is obtained by taking 8192 number of shots.}
\label{T3}
\end{figure}

Results of those experiments are shown in FIG.\ref{T3}.
For the generalization of two qubit case, we can consider a set of arbitrary two qubit orthogonal quantum states which is $\{|\phi_i\rangle\}=\{\alpha|00\rangle + \beta|01\rangle,\alpha|01\rangle + \beta|11\rangle, \beta|10\rangle-\alpha|11\rangle, \beta|00\rangle-\alpha|01\rangle \}$. Where, $\alpha$ and $\beta$ are real numbers satisfying, $|\alpha|^2+|\beta|^2=1$. This set is so chosen that the states are (a)orthogonal,
(b)not entangled, (c)different from Bell states, (d)do not
have definite parity and (e)contain single-superposed qubits (SSQB) (in this case second qubit is superposed). According to the theory(Sec.\ref{theory}), We can construct the $U_1$ and $U_2$ operators\cite{Chaung,Cross} for eignvalue array $\{1,1,-1,-1\}$ or $\{1,-1,1,-1\}$(Eqs.\ref{eq19},\ref{eq20}).
\begin{equation} 
U_1=\begin{bmatrix} Cos(\theta)&Sin(\theta)&0&0 \\ Sin(\theta)& -Cos(\theta)&0&0 \\ 0&0&Cos(\theta)&Sin(\theta) \\ 0&0&Sin(\theta)& -Cos(\theta) \end{bmatrix}
\label{eq19}
\end{equation}
\begin{equation} 
U_2=\begin{bmatrix} Cos(\theta)&Sin(\theta)&0&0 \\ Sin(\theta)& -Cos(\theta)&0&0 \\ 0&0&-Cos(\theta)&-Sin(\theta) \\ 0&0&-Sin(\theta)&Cos(\theta) \end{bmatrix}
\label{eq20}
\end{equation}
Where, 
\[\theta = 2 \times {Tan}^{-1}(\frac{\beta}{\alpha})\]
\section{Bell State Discrimination}
\label{bell}
\begin{figure}[H]
\includegraphics[scale=.3]{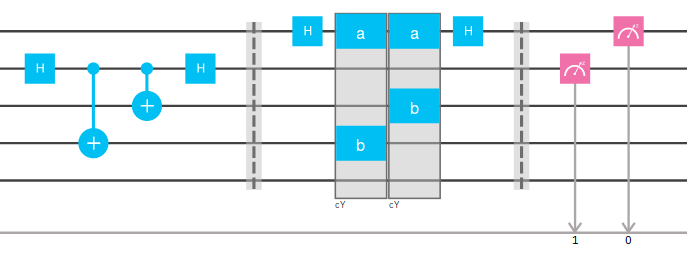}
\caption{The circuit for non-destructive discrimination of Bell-states(designed in Custom Topology). Here, first two qubits represent the ancilla qubit which are in state  $|00\rangle$. Last two qubit represent the quantum state. The meter represents a measurement of the qubit state in $\sigma_z$ basis.}
\label{BELL}
\end{figure}

 According to the theory(Sec.\ref{theory}), we can discriminate the Bell-states also. We showed  the Bell-states discrimination circuit in FIG.\ref{BELL}. The controlled-U operations\cite{Chaung,Cross} for this experiment are given in Eqs.\ref{eq17},\ref{eq18} .
\begin{equation} 
U_1=\begin{bmatrix} 0&0&0&1 \\ 0&0&1&0 \\ 0&1&0&0 \\ 1&0&0&0 \end{bmatrix} = \sigma_{x} \otimes \sigma_{x}
\label{eq17}
\end{equation}
\begin{equation} 
U_1=\begin{bmatrix} 0&0&0&-1 \\ 0&0&1&0 \\ 0&1&0&0 \\ -1&0&0&0 \end{bmatrix} = \sigma_{y} \otimes \sigma_{y}
\label{eq18}
\end{equation}

Therefore, the controlled-$U_1$ operation is CNOT$\otimes$CNOT and the controlled-$U_2$ operation is controlled-Y$\otimes$controlled-Y. Here, we prepare the state qubits using Pauli matrices, Hadamard gate and CONT gate sequencially. The schematic circuit diagram of the experiment is shown in the FIG.\ref{BELL}. The first part of the circuit diagram shoes the controlled-$U_1$ operation and secoend part shows the controlled-Y$\otimes$controlled-Y operation.

\begin{figure}[H]
\includegraphics[scale=.40]{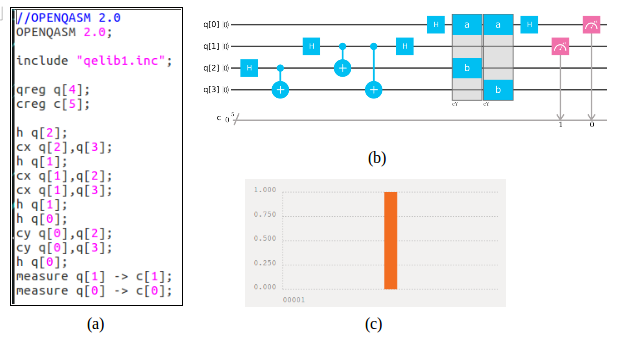}
\caption{Simulation-3: (a)Showes the code for the simulation of discrimination of $|\beta_1\rangle=\frac{1}{\sqrt{2}}(|00\rangle+|11\rangle)$ state.(b)Showes the circuit of the experiment. (c)Showes the simulation result of this experiment. Here, classical bit 0 and 1(cbit0 and cbit1) is the simulation result.}
\label{code3}
\end{figure}

\begin{figure}[H]
\includegraphics[scale=.40]{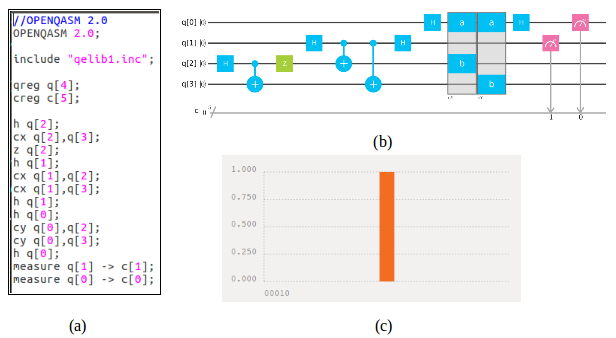}
\caption{Simulation-4: (a)Showes the code for the simulation of discrimination of $|\beta_2\rangle=\frac{1}{\sqrt{2}}(|00\rangle-|11\rangle)$ state.(b)Showes the circuit of the experiment. (c)Showes the simulation result of this experiment. Here, classical bit 0 and 1(cbit0 and cbit1) is the simulation result.}
\label{code4}
\end{figure}

\begin{figure}[H]
\includegraphics[scale=.40]{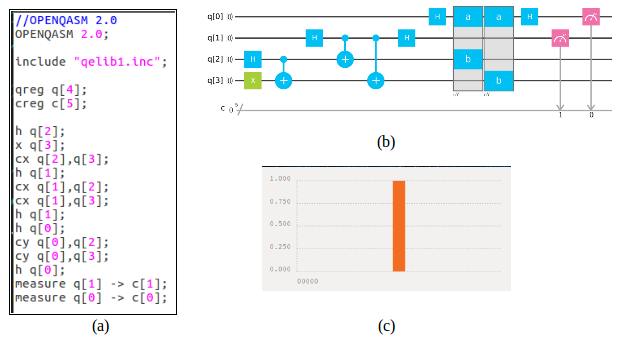}
\caption{Simulation-5: (a)Showes the code for the simulation of discrimination of $|\beta_3\rangle=\frac{1}{\sqrt{2}}(|01\rangle+|10\rangle)$ state.(b)Showes the circuit of the experiment. (c)Showes the simulation result of this experiment. Here, classical bit 0 and 1(cbit0 and cbit1) is the simulation result.}
\label{code5}
\end{figure}

\begin{figure}[H]
\includegraphics[scale=.40]{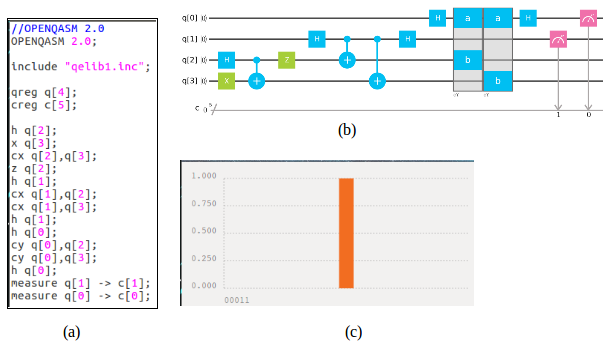}
\caption{Simulation-6: (a)Showes the code for the simulation of discrimination of $|\beta_4\rangle=\frac{1}{\sqrt{2}}(|01\rangle-|10\rangle)$ state.(b)Showes the circuit of the experiment. (c)Showes the simulation result of this experiment. Here, classical bit 0 and 1(cbit0 and cbit1) is the simulation result.}
\label{code6}
\end{figure}

We canot setup the Bell-state discrimination circuit in ibmqx2 and ibmqx4 quantum processors. In the particular case of 5-qubit IBM quantum computer, coupling is not present between all qubits(FIGs.\ref{transmon bowtie chip},\ref{ibmqx2}). So, in this paper we have done a simulation for four different bell states. We have shown the codes for discrimination of four different Bell-states and the corresponding results in FIGs.\ref{code3},\ref{code4},\ref{code5}  and \ref{code6}. Recently, Mitali Sisodia et al. publish a paper on non-destructive discrimination of Bell-states(\cite{Mitali}) using \textit{Panigrahi-circuit}. They discriminate the Bell-states using ibmqx2 processor with a high fidelity.  

\section{Some specifications about experimental setup}
\subsection{Specifications of ibmqx2 Processor}

\begin{figure}[H]
\begin{center}
\includegraphics[scale=.45]{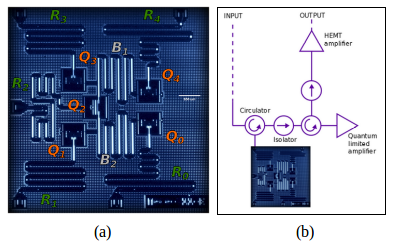}
\caption{(a)The ibmqx2 processor. (b)Schematic  diagram of the experimental setup.}
\label{a1}
\end{center}
\end{figure}

According to the IBMQX2:Sparrow\cite{IBM1}, the ibmqx2 processor went online 24th January, 2017. The connectivity is provided by two coplanar waveguide (CPW) resonators with resonances around 6.0 GHz (coupling Q2, Q3 and Q4) and 6.5 GHz (coupling Q0, Q1 and Q2). Each qubit has a dedicated CPW for control and readout. The FIG.\ref{a1} shows the chip layout and experimental setup. The fridge temperature of the setup is 0.0176 K.

\begin{figure}[H]
\includegraphics[scale=.35]{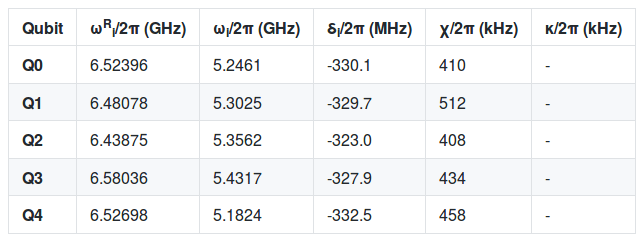}
\caption{The table shows some of the important experimental parameters for this device.}
\label{a2}
\end{figure}

where, $\omega^R_i$ is the resonance frequency of the resonator and $\omega_i = 2\pi\times(E_i - E_0)/h $ is the qubit frequency with $i=\{00001,00010,00100,01000,10000\}$. The anharmonicity $(\delta_i)$ is the difference between the frequency of the 1 to 2 transition and the 0 to 1 transition. That is, it is given by $\delta_i = 2\pi\times(E_{2i} - 2E_i+ E_0 )/h$. $\chi$ is the qubit-cavity coupling strength, and $\kappa$  is the cavity coupling to the environment.

\begin{figure}[H]
\includegraphics[scale=.37]{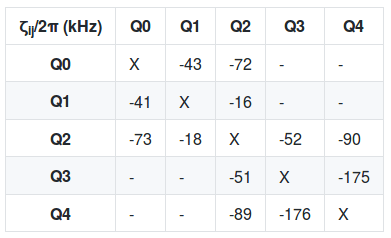}
\caption{In the crosstalk matrix, the error bar is less than 1 kHz for all $\zeta_{ij}$ and a dash indicates an interaction strength for that pair $< 25$ kHz.}
\label{a3}
\end{figure}

\begin{figure}[H]
\includegraphics[scale=.4]{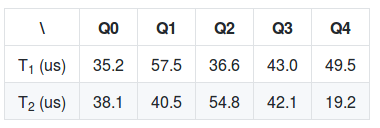}
\caption{The relaxation ($T_1$) and coherence ($T_2$) times for each qubit are given in this table. $T_2$ is measured with a \textit{Hahn echo} experiment. These values are averaged over 100 measurements each, spaced approximately by 12 hours, and performed between March and May 2017. The numbers in parentheses are standard errors of the mean.}
\label{a4}
\end{figure}

\subsection{Specifications of ibmqx4 Processor}

\begin{figure}[H]
\begin{center}
\includegraphics[scale=.38]{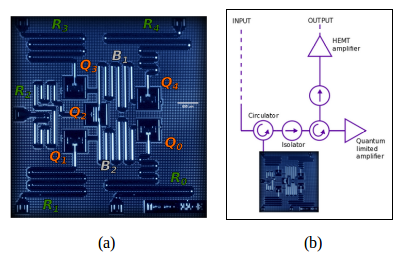}
\caption{(a)The ibmqx4 processor. (b)Schematic  diagram of the experimental setup.}
\label{b1}
\end{center}
\end{figure}

According to the IBMQX4:Raven\cite{IBM2}, the ibmqx4 processor went online 25th September, 2017. The connectivity is provided by two coplanar waveguide (CPW) resonators with resonances around 6.6 GHz (coupling Q2, Q3 and Q4) and 7.0 GHz (coupling Q0, Q1 and Q2). Each qubit has a dedicated CPW for control and readout.  The FIG.\ref{b1} shows the chip layout and experimental setup. The fridge temperature of the setup is 0.021 K.

\begin{figure}[H]
\includegraphics[scale=.40]{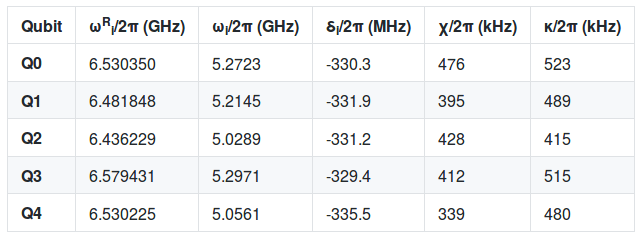}
\caption{The table shows some of the important experimental parameters for this device.}
\label{b2}
\end{figure}

where, $\omega^R_i$ is the resonance frequency of the resonator and $\omega_i = 2\pi\times(E_i - E_0)/h $ is the qubit frequency with $i=\{00001,00010,00100,01000,10000\}$. The anharmonicity $(\delta_i)$ is the difference between the frequency of the 1 to 2 transition and the 0 to 1 transition. That is, it is given by $\delta_i = 2\pi\times(E_{2i} - 2E_i+ E_0 )/h$. $\chi$ is the qubit-cavity coupling strength, and $\kappa$  is the cavity coupling to the environment.
\begin{figure}[H]
\includegraphics[scale=.48]{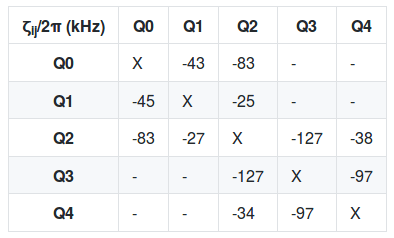}
\caption{In the crosstalk matrix, the error bar is less than 1 kHz for all $\zeta_{ij}$ and a dash indicates an interaction strength for that pair $< 25$ kHz.}
\label{b3}
\end{figure}
\begin{figure}[H]
\includegraphics[scale=.4]{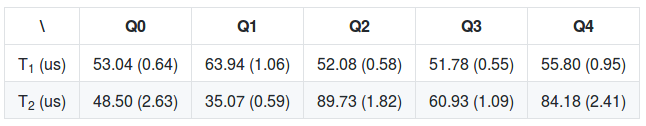}
\caption{The relaxation ($T_1$) and coherence ($T_2$) times for each qubit are given in the following table. $T_2$ is measured with a \textit{Hahn echo} experiment. These values are from single measurement on September 25, 2017. We will update these values when we have more statistics.}
\label{b4}
\end{figure}
\label{specifications}

\section{Conclusion}
\label{conclusion}
A general method for non-destructive discrimination of a set of orthogonal quantum states using quantum phase estimation algorithm has been descibed, and experimently implemented for a two
qubit case by NMR by V. S. Manu et al.\cite{Manu}. Here, we impliment the same experiment using two types of superconductivity based quantum processor(ibmqx2 and ibmqx4). As the direct measurements are performed only on the ancilla qubit, the discriminated states
are preserved. We also show the state-tomography for the single qubit experiment for both types of processor. The experiment confirmed that the arbitrary orthogonal quantum states can be discriminate in a non-destructive manner with a high fidelity. In the IBM quantum processors(ibmqx2 and ibmqx4), coupling is not present between all the qubits. Absence of couplings provides restriction on the applicability of CNOT gates. For this reason we cannot impliment the Bell-states discrimination circuits. For three qubit GHZ-states nad the Bell-states discrimination circuit can be done by using ibmqx3(16-qubit quantum processor). Some of the groups use this 16Q to impliment some experiments\cite{Wang,Ferrari}.
\section{Aknowledgement}
\label{aknowledgement}
A.M.  is  financially  supported  by  DST-Inspire Fellowship, Govt. of India. We thank Prof. Apoorva Patel of \textit{Indian Institute of Science, Bangalore} and Dr. K.V.R.M. Murali
member of \textit{Vijna Labs} for advise and encouragement.  We are extremely grateful to \textit{IBM research group} for the online accessibility of the real quantum processors(ibmqx2 and ibmqx4).


\begin{thebibliography}{20}%
\makeatletter
\providecommand \@ifxundefined [1]{%
 \@ifx{#1\undefined}
}%
\providecommand \@ifnum [1]{%
 \ifnum #1\expandafter \@firstoftwo
 \else \expandafter \@secondoftwo
 \fi
}%
\providecommand \@ifx [1]{%
 \ifx #1\expandafter \@firstoftwo
 \else \expandafter \@secondoftwo
 \fi
}%
\providecommand \natexlab [1]{#1}%
\providecommand \enquote  [1]{``#1''}%
\providecommand \bibnamefont  [1]{#1}%
\providecommand \bibfnamefont [1]{#1}%
\providecommand \citenamefont [1]{#1}%
\providecommand \href@noop [0]{\@secondoftwo}%
\providecommand \href [0]{\begingroup \@sanitize@url \@href}%
\providecommand \@href[1]{\@@startlink{#1}\@@href}%
\providecommand \@@href[1]{\endgroup#1\@@endlink}%
\providecommand \@sanitize@url [0]{\catcode `\\12\catcode `\$12\catcode
  `\&12\catcode `\#12\catcode `\^12\catcode `\_12\catcode `\%12\relax}%
\providecommand \@@startlink[1]{}%
\providecommand \@@endlink[0]{}%
\providecommand \url  [0]{\begingroup\@sanitize@url \@url }%
\providecommand \@url [1]{\endgroup\@href {#1}{\urlprefix }}%
\providecommand \urlprefix  [0]{URL }%
\providecommand \Eprint [0]{\href }%
\providecommand \doibase [0]{http://dx.doi.org/}%
\providecommand \selectlanguage [0]{\@gobble}%
\providecommand \bibinfo  [0]{\@secondoftwo}%
\providecommand \bibfield  [0]{\@secondoftwo}%
\providecommand \translation [1]{[#1]}%
\providecommand \BibitemOpen [0]{}%
\providecommand \bibitemStop [0]{}%
\providecommand \bibitemNoStop [0]{.\EOS\space}%
\providecommand \EOS [0]{\spacefactor3000\relax}%
\providecommand \BibitemShut  [1]{\csname bibitem#1\endcsname}%
\let\auto@bib@innerbib\@empty

\bibitem [{\citenamefont {Walgate}\ and\ \citenamefont {Short}(2000)}]{Walgate}%
  \BibitemOpen
  \bibfield  {author} {\bibinfo {author} {\bibfnamefont {J.}\ \bibnamefont
  {Walgate et al.,}}\ }\href {\doibase 85,4972(2000)} {\bibfield  {journal} {\bibinfo
   {journal} {Phys. Rev. Lett.}\ }\textbf {\bibinfo {volume} {85}}\ , \bibinfo {pages}
  {4972} (\bibinfo {year} {2000}).}\BibitemShut {Stop}%
  
\bibitem [{\citenamefont {Ghosh}\ and\ \citenamefont {}(2001)}]{Ghosh}%
  \BibitemOpen
  \bibfield  {author} {\bibinfo {author} {\bibfnamefont {S.}\ \bibnamefont
  {Ghosh et al.,}}\ }\href {\doibase 87,277902(2001)} {\bibfield  {journal} {\bibinfo
   {journal} {Phys. Rev. Lett.}\ }\textbf {\bibinfo {volume} {87}}\ , \bibinfo {pages}
  {277902} (\bibinfo {year} {2001}).}\BibitemShut {Stop}%
  
\bibitem [{\citenamefont {Virmani}\ and\ \citenamefont {}(2001)}]{Virmani}%
  \BibitemOpen
  \bibfield  {author} {\bibinfo {author} {\bibfnamefont {S.}\ \bibnamefont
  {Virmani et al.,}}\ }\href {\doibase 288,277902(2001)} {\bibfield  {journal} {\bibinfo
   {journal} {Phys. Lett. A}\ }\textbf {\bibinfo {volume} {288}}\ , \bibinfo {pages}
  {62} (\bibinfo {year} {2001}).}\BibitemShut {Stop}%
  
\bibitem [{\citenamefont {Chen}\ and\ \citenamefont {}(2001)}]{Chen}%
  \BibitemOpen
  \bibfield  {author} {\bibinfo {author} {\bibfnamefont {X. Y.}\ \bibnamefont
  {Chen et al.,}}\ }\href {\doibase 64,064303(2001)} {\bibfield  {journal} {\bibinfo
   {journal} {Phys. Rev. A}\ }\textbf {\bibinfo {volume} {64}}\ , \bibinfo {pages}
  {064303} (\bibinfo {year} {2001}).}\BibitemShut {Stop}%

\bibitem [{\citenamefont {Manu}\ and\ \citenamefont {Kumar}(2011)}]{Manu}%
  \BibitemOpen
  \bibfield  {author} {\bibinfo {author} {\bibfnamefont {V. S.}\ \bibnamefont
  {Manu}}\ and\ \bibinfo {author} {\bibfnamefont {A.}\ \bibnamefont
  {Kumar}},\ }\href {\doibase 10.1063/1.3635867} {\bibfield  {journal} {\bibinfo
   {journal} {AIP Conf. Proc.}\ }\textbf {\bibinfo {volume} {1384}},\ \bibinfo {pages}
  {229-240} (\bibinfo {year} {2011}).}\BibitemShut {Stop}%
  
\bibitem [{\citenamefont {Samal}\ \citenamefont {Gupta}\ \citenamefont {Panigrahi}\ and\ \citenamefont {Kumar}(2010)}]{Samal}%
  \BibitemOpen
  \bibfield  {author} {\bibinfo {author} {\bibfnamefont {J. R.}\ \bibnamefont
  {Samal et al.}}}, \href {\doibase } {\bibfield  {journal} {\bibinfo
   {journal} {Journal of Physics B: Atomic, Molecular and Optical
Physics}\ }\textbf {\bibinfo {volume} {43}},\ \bibinfo {pages}
  {095508} (\bibinfo {year} {2010}).}\BibitemShut {Stop}%

\bibitem [{\citenamefont {Majumder}\ \citenamefont {Mohapatra}\ and\ \citenamefont {Kumar}(2017)}]{Majumder}%
  \BibitemOpen
  \bibfield  {author} {\bibinfo {author} {\bibfnamefont {A.}\ \bibnamefont
  {Majumder et al.}}}, \href {\doibase arXiv:1707.07460 [quant-ph]} {\bibfield  {journal} {\bibinfo
   {journal} {arXiv:1707.07460 [quant-ph]}\ }\textbf {\bibinfo {volume} {}},\ \bibinfo {pages}
  {} (\bibinfo {year} {2017}).}\BibitemShut {Stop}%
    
\bibitem [{\citenamefont {Sisodia}\ \citenamefont {Shukla}\ and\ \citenamefont {Pathak}(2017)}]{Mitali}%
  \BibitemOpen
  \bibfield  {author} {\bibinfo {author} {\bibfnamefont {M.}\ \bibnamefont
  {Sisodia et al.}}}, \href {\doibase } {\bibfield  {journal} {\bibinfo
   {journal} {Phys. Lett. A}\ }\textbf {\bibinfo {volume} {381}},\ \bibinfo {pages}{3860-3874} (\bibinfo {year} {2017}).}\BibitemShut {Stop}%
   
\bibitem [{\citenamefont {Mandip}\ \citenamefont {}\ and\ \citenamefont {}(2015)}]{Mandip}%
  \BibitemOpen
  \bibfield  {author} {\bibinfo {author} {\bibfnamefont {M.}\ \bibnamefont
  {Singh}}}, \href {\doibase } {\bibfield  {journal} {\bibinfo
   {journal} {Phys. Lett. A}\ }, \textbf {\bibinfo {volume} {379}},\ \bibinfo {pages}{2001-2006} (\bibinfo {year} {2015}).}\BibitemShut {Stop}%

\bibitem [{\citenamefont {Alsina}\ and\ \citenamefont {}(2016)}]{Alsina}%
  \BibitemOpen
  \bibfield  {author} {\bibinfo {author} {\bibfnamefont {D.}\ \bibnamefont
  {Alsina et al.,}}\ }\href {\doibase 94,012314(2016)} {\bibfield  {journal} {\bibinfo
   {journal} {Phys. Rev. A}\ },\textbf {\bibinfo {volume} {94}}\ , \bibinfo {pages}
  {012314} (\bibinfo {year} {2016}).}\BibitemShut {Stop}%
  
\bibitem [{\citenamefont {Beltra}\ and\ \citenamefont {}(2016)}]{Beltra}%
  \BibitemOpen
  \bibfield  {author} {\bibinfo {author} {\bibfnamefont {R. L.}\ \bibnamefont
  {Beltra et al.,}}\ }\href {\doibase 10.3390(2016)} {\bibfield  {journal} {\bibinfo
   {journal} {MDPI(Computers)}\ },\textbf {\bibinfo {volume} {5}}\ , \bibinfo {pages}
  {24} (\bibinfo {year} {2016}).}\BibitemShut {Stop}%
  
\bibitem [{\citenamefont {Du}\ and\ \citenamefont {}(2010)}]{Du}%
  \BibitemOpen
  \bibfield  {author} {\bibinfo {author} {\bibfnamefont {J.}\ \bibnamefont
  {Du et al.,}}\ }\href {\doibase } {\bibfield  {journal} {\bibinfo
   {journal} {Phys. Rev. Lett.}\ },\textbf {\bibinfo {volume} {104}}\ , \bibinfo {pages}
  {030502} (\bibinfo {year} {2010}).}\BibitemShut {Stop}%
  
\bibitem [{\citenamefont {Guzik}\ and\ \citenamefont {}(2005)}]{Guzik}%
  \BibitemOpen
  \bibfield  {author} {\bibinfo {author} {\bibfnamefont {A. S.}\ \bibnamefont
  {Guzik et al.,}}\ }\href {\doibase 10.1126/science.1113479} {\bibfield  {journal} {\bibinfo
   {journal} {SCIENCE}\ },\textbf {\bibinfo {volume} {309}}\ , \bibinfo {pages}
  {1704-1707} (\bibinfo {year} {2005}).}\BibitemShut {Stop}%
  
\bibitem [{\citenamefont {You}\ and\ \citenamefont {}(2011)}]{You}%
  \BibitemOpen
  \bibfield  {author} {\bibinfo {author} {\bibfnamefont {J. Q.}\ \bibnamefont
  {You et al.,}}\ }\href {\doibase 10.1038/nature10122 } {\bibfield  {journal} {\bibinfo
   {journal} {NATURE}\ },\textbf {\bibinfo {volume} {474}}\ , \bibinfo {pages}
  {589-597} (\bibinfo {year} {2011}).}\BibitemShut {Stop}%

\bibitem [{\citenamefont {Clarke}\ and\ \citenamefont {}(2008)}]{Clarke}%
  \BibitemOpen
  \bibfield  {author} {\bibinfo {author} {\bibfnamefont {J.}\ \bibnamefont
  {Clarke et al.,}}\ }\href {\doibase 10.1038/nature07128 } {\bibfield  {journal} {\bibinfo
   {journal} {NATURE}\ },\textbf {\bibinfo {volume} {453}}\ , \bibinfo {pages}
  {1031-1042} (\bibinfo {year} {2008}).}\BibitemShut {Stop}%
  
\bibitem [{\citenamefont {Shi}\ and\ \citenamefont {}(2016)}]{Shi}%
  \BibitemOpen
  \bibfield  {author} {\bibinfo {author} {\bibfnamefont {Run-hau}\ \bibnamefont
  {Shi et al.,}}\ }\href {\doibase 10.1038/srep19655 } {\bibfield  {journal} {\bibinfo
   {journal} {SCIENTIFIC REPORTS}\ },\textbf {\bibinfo {volume} {6}}\ , \bibinfo {pages}
  {19655} (\bibinfo {year} {2016}).}\BibitemShut {Stop}%
  
\bibitem [{\citenamefont {Rosenberg}\ and\ \citenamefont {}(2017)}]{Rosenberg}%
  \BibitemOpen
  \bibfield  {author} {\bibinfo {author} {\bibfnamefont {D.}\ \bibnamefont
  {Rosenberg et al.,}}\ }\href {\doibase 1706.04116v1[quant-ph]13Jun2017} {\bibfield  {journal} {\bibinfo
   {journal} {arXiv:1706.04116v1 [quant-ph] 13 Jun 2017}\ }, \bibinfo {pages}
  {} (\bibinfo {year} {2017}).}\BibitemShut {Stop}%
  
\bibitem [{\citenamefont {Behera}\ and\ \citenamefont {}(2017)}]{Behera}%
  \BibitemOpen
  \bibfield  {author} {\bibinfo {author} {\bibfnamefont {B. K.}\ \bibnamefont
  {Behera et al.,}}\ }\href {\doibase 1707.00182v2[quant-ph]10July2017} {\bibfield  {journal} {\bibinfo{journal} {arXiv:1707.00182v2 [quant-ph] 10 July 2017}\ }, \bibinfo {pages}
  {} (\bibinfo {year} {2017}).}\BibitemShut {Stop}%
  
\bibitem [{\citenamefont {Kalra}\ and\ \citenamefont {}(2017)}]{Kalra}%
  \BibitemOpen
  \bibfield  {author} {\bibinfo {author} {\bibfnamefont {A. R.}\ \bibnamefont
  {Kalra et al.,}}\ }\href {\doibase 1707.09462v1[quant-ph]29June2017} {\bibfield  {journal} {\bibinfo{journal} {arXiv:1707.09462v1 [quant-ph] 29 June 2017}\ }, \bibinfo {pages}
  {} (\bibinfo {year} {2017}).}\BibitemShut {Stop}%
  
\bibitem [{\citenamefont {Ma}\ and\ \citenamefont {}(2017)}]{Ma}%
  \BibitemOpen
  \bibfield  {author} {\bibinfo {author} {\bibfnamefont {Shi-Yuan}\ \bibnamefont
  { Ma  et al.,}}\ }\href {\doibase 10.18576/amis/110531} {\bibfield  {journal} {\bibinfo
   {journal} {Appl. Math. Inf. Sci.}\ }, \textbf {\bibinfo {volume} {11}}\ ,\bibinfo {pages}
  {1519-1526} (\bibinfo {year} {2017}).}\BibitemShut {Stop}%
  
 \bibitem [{\citenamefont {Devitt}\ and\ \citenamefont {}(2016)}]{Devitt}%
  \BibitemOpen
  \bibfield  {author} {\bibinfo {author} {\bibfnamefont {S. J.}\ \bibnamefont
  {Devitt,}}\ }\href {\doibase 10.1103/PhysRevA.94.032329} {\bibfield  {journal} {\bibinfo
   {journal} {Phys. Rev. A}\ },\textbf {\bibinfo {volume} {94}}\ , \bibinfo {pages}
  {032329} (\bibinfo {year} {2016}).}\BibitemShut {Stop}%
  
\bibitem [{\citenamefont {Cross}\ and\ \citenamefont {}(2016)}]{Cross}%
  \BibitemOpen
  \bibfield  {author} {\bibinfo {author} {\bibfnamefont {A. W.}\ \bibnamefont
  {Cross et al.,}}\ }\href {\doibase} {\bibfield  {journal} {\bibinfo
   {journal} {Open Quantum Assembly Language}\ },\textbf {\bibinfo {volume} {}}\  \bibinfo {pages}
  {} (\bibinfo {year} {2017}).}\BibitemShut {Stop}%
  
\bibitem [{\citenamefont {Nielsen}\ and\ \citenamefont {Chaung}(2010)}]{Chaung}%
  \BibitemOpen
  \bibfield  {author} {\bibinfo {author} {\bibfnamefont {M. A.}\ \bibnamefont
  {Nielsen et al.,}}\ }\href {\doibase} {\bibfield  {journal} {\bibinfo
   {journal}{Quantum Computation and Quantum Information}\ },\textbf {\bibinfo {volume} {10th Anniversary Edition, Cambridge University Press}}\ ,\bibinfo {pages}
  {ISBN-13 978-1-107-61919-7} (\bibinfo {year} {2010}).}\BibitemShut {Stop}%


\bibitem [{\citenamefont {Yuanhao Wang}\ and\ \citenamefont {}(2018)}]{Wang}%
  \BibitemOpen
  \bibfield  {author} {\bibinfo {author} {\bibfnamefont {Yuanhao}\ \bibnamefont
  {Wang et al.,}}\ }\href {\doibase arXiv:1801.03782v1 [quant-ph]} {\bibfield  {journal} {\bibinfo{journal} {arXiv:1801.03782v1 [quant-ph] 11 January 2018}\ }, \bibinfo {pages}
  {} (\bibinfo {year} {2018}).}\BibitemShut {Stop}%

\bibitem [{\citenamefont {Davide Ferrari}\ and\ \citenamefont {}(2018)}]{Ferrari}%
  \BibitemOpen
  \bibfield  {author} {\bibinfo {author} {\bibfnamefont {D.}\ \bibnamefont
  {Ferrari et al.,}}\ }\href {\doibase arXiv:1801.02363v1 [quant-ph]} {\bibfield  {journal} {\bibinfo{journal} {arXiv:1801.02363v1 [quant-ph] 8 January 2018}\ }, \bibinfo {pages}
  {} (\bibinfo {year} {2018}).}\BibitemShut {Stop}%

\bibitem [{\citenamefont {IBM1}\ and\ \citenamefont {}()}]{IBM1}%
  \BibitemOpen
  \bibfield  {author} {\bibinfo {author} {\bibfnamefont {IBM QX2: Sparrow}\ \bibnamefont
  {}}\ }\href {\doibase } {\bibfield  {journal} {\bibinfo{journal} {https://github.com/QISKit/ibmqx-backend-information/blob/master/backends/ibmqx2/README.md}\ } \bibinfo {pages}
  {} (\bibinfo {year} {2017}).}\BibitemShut {Stop}%

\bibitem [{\citenamefont {IBM2}\ and\ \citenamefont {}()}]{IBM2}%
  \BibitemOpen
  \bibfield  {author} {\bibinfo {author} {\bibfnamefont {IBM QX4: Raven}\ \bibnamefont
  {}}\ }\href {\doibase } {\bibfield  {journal} {\bibinfo{journal} {https://github.com/QISKit/ibmqx-backend-information/blob/master/backends/ibmqx4/README.md}\ } \bibinfo {pages}
  {} (\bibinfo {year} {2017}).}\BibitemShut {Stop}%
\end{thebibliography}
\end{document}